# Tight Lower Bounds for Query Processing on Streaming and External Memory Data

Martin Grohe[1], Christoph Koch[2], and Nicole Schweikardt[1]

[1] Institut für Informatik, Humboldt-Universität Berlin, Germany
{grohe,schweika}@informatik.hu-berlin.de
[2] Database Group, Universität des Saarlandes, Saarbrücken, Germany
koch@cs.uni-sb.de

**Abstract.** We study a clean machine model for external memory and stream processing. We show that the number of scans of the external data induces a strict hierarchy (as long as work space is sufficiently small, e.g., polylogarithmic in the size of the input). We also show that neither joins nor sorting are feasible if the product of the number $r(n)$ of scans of the external memory and the size $s(n)$ of the internal memory buffers is sufficiently small, e.g., of size $o(\sqrt[5]{n})$. We also establish tight bounds for the complexity of XPath evaluation and filtering.

## 1 Introduction

It is generally assumed that databases have to reside in external, inexpensive storage because of their sheer size. Current technology for external storage systems (disks and tapes) presents us with a reality that performance-wise, a small number of *sequential scans* of the data is strictly preferable over random data accesses. Indeed, the combined latencies and access times of moving to a certain position in external storage are by orders of magnitude greater than actually reading a small amount of data once the read head has been placed on its starting position.

Database engines rely on main memory buffers for assuring acceptable performance. These are usually small compared to the size of the externally stored data. Database technology – in particular query processing technology – has developed around this notion of memory hierarchies with layers of greatly varying sizes and access times. There has been a wealth of research on query processing and optimization along these lines (cf. e.g. [27, 14, 32, 22]). It seems that the current technologies scale up to current user expectations, but on closer investigation it may appear that our theoretical understanding of the problems involved – and of optimal algorithms for these problems – is not quite as developed.

Recently, data stream processing has become an object of study by the data management community (e.g. [15]) but from the viewpoint of database theory, this is, in fact, a special case of the query processing problem on data in external storage where we are limited to a single scan of the input data.

In summary, it appears that there are a variety of data management and query processing problems in which a comparably small but efficiently accessible main memory buffer is available and where accessing external data is costly and is best performed by sequential read/write scans. This calls for an appropriate formal model that captures the essence of external memory and stream processing. In this paper, we study such a model, which employs a Turing machine with one *external memory tape* (external tape for short) and a number *internal memory tapes* (internal tapes for short). The external tape initially holds the input; the internal tapes correspond to the main memory buffers of a database management system and are thus usually small compared to the input.

As computational resources for inputs of size $n$, we study the space $s(n)$ available on the internal tapes and the number $r(n)$ of scans of (or, random accesses to) the external tape, and we write $ST(r, s)$ to denote the class of all problems solvable by $(r, s)$-bounded Turing machines, i.e., Turing machines which comply to the resource bounds $r(n)$ and $s(n)$ on inputs of size $n$.

Formally, we model the number of scans, respectively the number of random accesses, by the number of reversals of the Turing machine's read/write head on the external tape. The number of reversals of the read/write head on the internal tapes remains unbounded. The reversals done by a read/write head are a clean and fundamental notion [8], but of course real external storage technology based on disks does not allow to reverse their direction of rotation. On the other hand, we can of course simulate $k$ forward scans by $2k$ reversals in our machine model — and allowing for forward as well as backward scans makes our *lower* bound results even stronger.

As we allow the external tape to be both read and written to, the external tape can be viewed, for example, as modeling a hard disk. By closely watching reversals of the external tape head, anything close to random I/O will result in a very considerable number of reversals, while a full sequential scan of the external data can be effected cheaply. We will obtain strong lower bounds in this paper that show that even if the external tape (whose size we do not put a bound on) may be written to and re-read, certain bounds cannot be improved upon. For our matching upper bounds, we will usually not write to the external tape. Whenever one of our results requires writing to the external tape, we will explicitly indicate this.

The model is similar in spirit to the frameworks used in [18, 19], but differs from the previously considered *reversal complexity* framework [8]. Reversal complexity is based on Turing machines with a single read/write tape and the overall number of reversals of the read/write head the main computational resource. In our notion, only the number of reversals on the external tape is bounded, while reversals on the internal tapes are free; however, the space on the internal tapes is considered to be a limited resource.[3]

---

[3] The justification for this assumption is simply that accessing data on disks is currently about five to six orders of magnitude slower than accessing main memory. For that reason, processor cycles and main memory access times are often neglected when estimating query cost in relational query optimizers, where cost measures are often exclusively based on the amount of expected page I/O as well as disk latency and access times. Moreover, by taking buffer space rather than running time as a parameter, we obtain more robust complexity classes that rely less on details of the machine model (see also [31]).

Apart from formalizing the $ST(r,s)$ model, we study its properties and locate a number of data management problems in the hierarchy of $ST(\cdot,\cdot)$ classes. Our technical contributions are as follows:

- We prove a reduction lemma (Lemma 4.1) which allows easy lower bound proofs for certain problems.
- We prove a hierarchy (Corollary 4.11 and Theorem 4.10), stating for each fixed number $k$ that $k+1$ scans of the external memory tape are strictly more powerful than $k$ scans of the external memory tape.
- We consider machines where the product of the number of scans of the external memory tape, $r(n)$, and internal memory tape size, $s(n)$, is of size $o\!\left(\frac{n}{\log n}\right)$, where $n$ is the input size, and show that *joins* cannot be computed by $(r,s)$-bounded Turing machines (cf., Lemma 4.4).
- We show that the *sorting* problem cannot be solved with $(o(\sqrt[5]{n}), O(\sqrt[5]{n}))$-bounded Turing machines that are not allowed to write intermediate results to the external memory tape (cf., Corollary 4.9).
- We show (cf., Theorem 4.5) that for some *XQuery* queries, filtering is impossible for machines with $r(T) \cdot s(T) \in o\!\left(\frac{n}{\log n}\right)$, where $n$ is the size of the input XML document $T$.
- We show (cf., Corollary 5.5) that for some *Core XPath* [12] queries, filtering is impossible for machines with $r(T) \cdot s(T) \in o(d)$, where $d$ denotes the *depth* of the input XML document $T$. Furthermore, we show that the lower bound on *Core XPath* is *tight* in that we give an algorithm that solves the Core XPath filtering problem with a single scan of the external data (zero reversals) and $O(d)$ buffer space.

The primary technical machinery that we use for obtaining lower bounds is that of *communication complexity* (cf. [21]). Techniques from communication complexity have been used previously to study queries on streams [4, 6, 2, 3, 5, 23, 24, 18]. The work reported on in [4] addresses the problem of determining whether a given relational query can be evaluated scalably on a data stream or not at all. In comparison, we ask for tight bounds on query evaluation problems, i.e. we give algorithms for query evaluation that are in a sense worst-case optimal. As we do, the authors of [6] study XPath evaluation; however, they focus on *instance data complexity* while we study worst-case bounds. This allows us to find strong and tight bounds for a greater variety of query evaluation problems. Many of our results apply beyond stream processing in a narrow sense to a more general framework of queries on data in external storage. Also, our worst-case bounds apply for *any* evaluation algorithm possible, that is, our bounds are not in terms of complexity classes closed under reductions that allow for nonlinear expansions of the input (such as LOGSPACE) as is the case for the work on the complexity of XPath in [12, 13, 28].

Lower bound results for a machine model with *multiple* external memory tapes (or harddisks) are presented in [17]. In the present paper, we only consider a single external memory tape, and are consequently able to show (sometimes exponentially) stronger lower bounds.

The present paper is the full version of the conference contribution [16].

## 2   Preliminaries

In this section we fix some basic notation concerning trees, streams, and query languages. We write $\mathbb{N}$ for the set of non-negative integers. If $M$ is a set, then $2^M$ denotes the set of all subsets of $M$. Throughout this paper we make the following convention: Whenever the letters $r$ and $s$ denote functions from $\mathbb{N}$ to $\mathbb{N}$, then these functions are *monotone*, i.e., we have $r(x) \leqslant r(y)$ and $s(x) \leqslant s(y)$ for all $x, y \in \mathbb{N}$ with $x \leqslant y$.

**Trees and Streams.** We use standard notation for trees and streamed trees (i.e. *documents*). In particular, we write $Doc(T)$ to denote the XML document associated with an XML document tree $T$. An example is given in Figure 1. Some more details on trees and streams can be found in Appendix A.

**Query Languages.** By $Eval(\cdot, \cdot)$ we denote the evaluation function that maps each tuple $(Q, T)$, consisting of a query $Q$ and a tree $T$ to the corresponding query result. Let $\mathcal{Q}$ be a query language and let $\mathcal{T}_1 \subseteq Trees_\tau$ and $\mathcal{T}_2 \subseteq \mathcal{T}_1$. We say that $\mathcal{T}_2$ *can be filtered from $\mathcal{T}_1$ by a $\mathcal{Q}$-query* if, and only if, there is a query $Q \in \mathcal{Q}$ such that the following is true for all $T \in \mathcal{T}_1$: $T \in \mathcal{T}_2 \iff Eval(Q, T) \neq \emptyset$.

We assume that the reader is familiar with first-order logic (*FO*) and monadic second-order logic (*MSO*). An *FO-* or *MSO-sentence* (i.e., a formula without any free variable) specifies a Boolean query, whereas a formula with exactly one free first-order variable specifies a unary query, i.e., a query which selects a set of nodes from the underlying input tree.

It is well-known [9, 30] that the MSO-definable Boolean queries on binary trees are exactly the (Boolean) queries that can be defined by finite (deterministic or nondeterministic) bottom-up tree automata. An analogous statement is true about MSO on unranked trees and unranked tree automata [7].

Theorem 4.5 in section 4 gives a lower bound on the worst case complexity of the language *XQuery*. As we prove a lower bound for *one* particular XQuery query, we do not give a formal definition of the language but refer to [33].

Apart from *FO*, *MSO*, and XQuery, we also consider a fragment of the XPath language, Core XPath [12, 13]. As we will prove not only lower, but also upper bounds for Core XPath, we give a precise definition of this query language in Appendix B. An example of a Core XPath query is

/descendant::∗[child::A and child::B]/child::∗,

which selects all children of descendants of the root node that (i.e., the descendants) have a child node labeled A and a child node labeled B.

Core XPath is a strict fragment of XPath [12], both syntactically and semantically. It is known that Core XPath is in LOGSPACE w.r.t. data complexity and P-complete w.r.t. combined complexity [13]. In [12], it is shown that Core XPath can be evaluated in time $O(|Q| \cdot |D|)$, where $|Q|$ is the size of the query and $|D|$ is the size of the XML data. Furthermore, every Core XPath query is equivalent to a unary MSO query on trees (cf., e.g., [11]).

**Communication complexity.** To prove basic properties and lower bounds for our machine model, we use some notions and results from *communication complexity*, cf., e.g., [21].

Let $A, B, C$ be sets and let $F : A \times B \to C$ be a function. In Yao's [34] basic model of communication two players, Alice and Bob, jointly want to evaluate $F(x,y)$, for input values $x \in A$ and $y \in B$, where Alice only knows $x$ and Bob only knows $y$. The two players can exchange messages according to some fixed protocol $\mathcal{P}$ that depends on $F$, but not on the particular input values $x, y$. The exchange of messages starts with Alice sending a message to Bob and ends as soon as one of the players has enough information on $x$ and $y$ to compute $F(x,y)$.

$\mathcal{P}$ is called a *k-round protocol*, for some $k \in \mathbb{N}$, if the exchange of messages consists, for each input $(x,y) \in A \times B$, of at most $k$ rounds. The *cost* of $\mathcal{P}$ on input $(x,y)$ is the number of bits communicated by $\mathcal{P}$ on input $(x,y)$. The *cost of* $\mathcal{P}$ is the *maximal* cost of $\mathcal{P}$ over all inputs $(x,y) \in A \times B$. The *communication complexity of* $F$, *comm-compl*$(F)$, is defined as the minimum cost of $\mathcal{P}$, over all protocols $\mathcal{P}$ that compute $F$. For $k \geqslant 1$, the *k-round communication complexity of* $F$, *comm-compl*$_k(F)$, is defined as the minimum cost of $\mathcal{P}$, over all $k$-round protocols $\mathcal{P}$ that compute $F$.

Many powerful tools are known for proving lower bounds on communication complexity, cf., e.g., [21]. In the present paper we will use the following basic lower bound for the problem of deciding whether two sets are disjoint.

**Definition 2.1.** For $n \in \mathbb{N}$ let the function $Disj_n : 2^{\{1,\ldots,n\}} \times 2^{\{1,\ldots,n\}} \to \{0,1\}$ be given via

$$Disj_n(X,Y) := \begin{cases} 1 \ , & \text{if } X \cap Y = \emptyset \\ 0 \ , & \text{otherwise.} \end{cases}$$

$\square$

**Theorem 2.2 (cf., e.g., [21]).** *For every* $n \in \mathbb{N}$, *comm-compl*$(Disj_n) \geqslant n$.

## 3 Machine Model

We consider Turing machines with (1) an input tape, which is a read/write tape and will henceforth be called "external memory tape" or "external tape", for short, (2) an arbitrary number $u$ of work tapes, which will henceforth be called "internal memory tapes" or "internal tapes", for short, and, if needed, (3) an additional write-only output tape.

Let $M$ be such a Turing machine and let $\rho$ be a run of $M$. By rev$(\rho)$ we denote the number of times the external memory tape's head changes its direction in the run $\rho$. For $i \in \{1, \ldots, u\}$ we let space$(\rho, i)$ be the number of cells of internal memory tape $i$ that are used by $\rho$.

**The class $\boldsymbol{ST}(r,s)$ for strings.**

**Definition 3.1 ($\boldsymbol{ST}(r,s)$ for strings).** Let $r : \mathbb{N} \to \mathbb{N}$ and $s : \mathbb{N} \to \mathbb{N}$.

(a) A Turing machine $M$ is $(r,s)$-*bounded*, if every run $\rho$ of $M$ on an input of length $n$ satisfies the following conditions:

(1) $\rho$ is finite,   (2) $1 + \mathrm{rev}(\rho) \le r(n)$,[4]  and   (3) $\sum_{i=1}^{u} \mathrm{space}(\rho, i) \le s(n)$, where $u$ is the number of internal tapes of $M$.

(b) A string-language $L \subseteq \Sigma^*$ belongs to the class $ST(r, s)$ (resp., $NST(r, s)$), if there is a deterministic (respectively, nondeterministic) $(r, s)$-bounded Turing machine which accepts exactly those $w \in \Sigma^*$ that belong to $L$.

(c) A function $f : \Sigma^* \to \Sigma^*$ belongs to the class $ST(r, s)$, if there is a deterministic $(r, s)$-bounded Turing machine which produces, for each input string $w \in \Sigma^*$, the string $f(w)$ on its write-only output tape.               □

For classes $R$ and $S$ of functions, we let   $ST(R, S) := \bigcup_{r \in R, s \in S} ST(r, s)$.
If $k \in \mathbb{N}$ is a constant, then we write $ST(k, s)$ instead of $ST(r, s)$, where $r$ is the function with $r(x) = k$ for all $x \in \mathbb{N}$. We freely combine these notations and use them for $NST(\cdot, \cdot)$ instead of $ST(\cdot, \cdot)$, too.

If we think of the external memory tape of an $(r, s)$-bounded Turing machine as representing the incoming stream, stored on a hard disk, then admitting the external memory tape's head to reverse its direction might not be very realistic. But as we mainly use our model to prove lower bounds, it does not do any harm either, since the reversals can be used to simulate *random access*. Random access can be introduced explicitly into our model as follows: A *random access Turing machine* is a Turing machine $M$ which has a special internal memory tape that is used as *random access address tape*, i.e., on which only binary strings can be written. Such a binary string is interpreted as a positive integer specifying an external memory address, that is, the position index number of a cell on the external tape (we think of the external tape cells being numbered by positive integers). The machine has a special state $q_{\mathrm{ra}}$. If $q_{\mathrm{ra}}$ is entered, then in one step the external memory tape head is moved to the cell that is specified by the number on the random access address tape, and the content of the random access address tape is deleted.

**Definition 3.2.** Let $q, r, s : \mathbb{N} \to \mathbb{N}$. A random access Turing machine $M$ is $(q, r, s)$-*bounded*, if it is $(r, s)$-bounded (in the sense of an ordinary Turing machine) and, in addition, every run $\rho$ of $M$ on an input of length $n$ involves at most $q(n)$ random accesses.               □

Noting that a random access can be simulated with at most 2 changes of the direction of the external memory tape head, one immediately obtains:

**Lemma 3.3.** *Let $q, r, s : \mathbb{N} \to \mathbb{N}$. If a problem can be solved by a $(q, r, s)$-bounded random access Turing machine, then it can also be solved by an $(r + 2q, O(s))$-bounded Turing machine.*

In the subsequent parts of this paper, we will concentrate on ordinary Turing machines (without random access). Via Lemma 3.3, all results can be transferred from ordinary Turing machines to random access Turing machines.

---

[4] It is convenient for technical reasons to add 1 to the number $\mathrm{rev}(\rho)$ of changes of the head direction. As defined here, $r(n)$ bounds the number of sequential scans of the external memory tape rather than the number of changes of head directions.

**The class $ST(r, s)$ for trees.** We make an analogous definition to $ST(r, s)$ on strings for trees. This definition is given in detail in Appendix C.

## 4   Lower bounds for the ST model

**A reduction lemma.** The following lemma provides a convenient tool for showing that a problem $L$ does not belong to $ST(r, s)$. The lemma's assumption can be viewed as a *reduction* from the problem $Disj_n(\cdot, \cdot)$ to the problem $L$.

**Lemma 4.1.** *Let $\Sigma$ be an alphabet and let $\lambda : \mathbb{N} \to \mathbb{N}$ such that the following is true: For every $n_0 \in \mathbb{N}$ there is an $n \geqslant n_0$ and functions $f_n, g_n : 2^{\{1,\dots,n\}} \to \Sigma^*$ such that for all $X, Y \subseteq \{1, \dots, n\}$ the string $f_n(X)g_n(Y)$ has length $\leqslant \lambda(n)$.*

*Then we have for all $r, s : \mathbb{N} \to \mathbb{N}$ with $r(\lambda(n)) \cdot s(\lambda(n)) \in o(n)$, that there is no $(r, s)$-bounded deterministic Turing machine which accepts a string of the form $f_n(X)g_n(Y)$ if, and only if, $X \cap Y = \emptyset$.*

The proof of this lemma can be found in Appendix D.

**Disjointness.** Every $n$-bit string $x = x_1 \cdots x_n \in \{0, 1\}^n$ specifies a set $S(x) := \{i : x_i = 1\} \subseteq \{1, \dots, n\}$. Let $L_{Disj}$ consist of those strings $x\#y$ where $x$ and $y$ specify disjoint subsets of $\{1, \dots, n\}$, for some $n \geqslant 1$. That is,

$$L_{Disj} \quad := \quad \big\{ \ x\#y \ : \ \text{ex. } n \geqslant 1 \text{ with } x, y \in \{0, 1\}^n \text{ and } S(x) \cap S(y) = \emptyset \ \big\}.$$

From Lemma 4.1 one easily obtains

**Proposition 4.2.** *Let $r : \mathbb{N} \to \mathbb{N}$ and $s : \mathbb{N} \to \mathbb{N}$. If $r(n) \cdot s(n) \in o(n)$, then $L_{Disj} \notin ST(r, s)$.*

The proof can be found in Appendix E. The bound given by Proposition 4.2 is tight, as it can be easily seen that $L_{Disj} \in ST(r, s)$ for all $r, s : \mathbb{N} \to \mathbb{N}$ with $r(n) \cdot s(n) \in \Omega(n)$.

**Joins.** Let $\tau$ be the set of tag names { rels, rel1, rel2, tuple, no1, no2, 0, 1 }. We represent a pair $(A, B)$ of finite relations $A, B \subseteq \mathbb{N}^2$ as a $\tau$-tree $T(A, B)$ whose associated XML document $Doc(T(A, B))$ is a $\Sigma_\tau$-string of the following form: For each number $i \in \mathbb{N}$ let $Bin(i) = b_{\ell_i}^{(i)} \cdots b_0^{(i)}$ be the binary representation of $i$. For each tuple $(i, j) \in \{1, \dots, n\}^2$ let $Doc(i, j) :=$

$$\langle\text{tuple}\rangle \ \langle\text{no1}\rangle \ \langle b_{\ell_i}^{(i)}/\rangle \ \cdots \ \langle b_0^{(i)}/\rangle \ \langle/\text{no1}\rangle \ \langle\text{no2}\rangle \ \langle b_{\ell_j}^{(j)}/\rangle \ \cdots \ \langle b_0^{(j)}/\rangle \ \langle/\text{no2}\rangle \ \langle/\text{tuple}\rangle \ .$$

For each finite relation $A \subseteq \mathbb{N}^2$ let $t_1, \dots, t_{|A|}$ be the lexicographically ordered list of all tuples in $A$. We let $Doc(A) := Doc(t_1) \ \cdots \ Doc(t_{|A|})$. Finally, we let

$$Doc(T(A, B)) \quad := \quad \langle\text{rels}\rangle \ \langle\text{rel1}\rangle \ Doc(A) \ \langle/\text{rel1}\rangle \ \langle\text{rel2}\rangle \ Doc(B) \ \langle/\text{rel2}\rangle \ \langle/\text{rels}\rangle.$$

It is straightforward to see that the string $Doc(T(A, B))$ has length $O\big((|A| + |B|) \cdot \log n\big)$, if $A, B \subseteq \{1, \dots, n\}^2$.

We write  $A \bowtie_1 B$  to denote the join of $A$ and $B$ on their first component, i.e., $A \bowtie_1 B := \{\, (x,y) \, : \, \exists z \; A(z,x) \wedge B(z,y) \,\}$. We let

$$\mathcal{T}_{Rels} := \{\, T(A,B) \, : \, A,B \subseteq \mathbb{N}^2, \; A,B \text{ finite}\,\}$$
$$\mathcal{T}_{EmptyJoin} := \{\, T(A,B) \in \mathcal{T}_{Rels} \, : \, A \bowtie_1 B = \emptyset \,\}$$
$$\mathcal{T}_{NonEmptyJoin} := \{\, T(A,B) \in \mathcal{T}_{Rels} \, : \, A \bowtie_1 B \neq \emptyset \,\}.$$

**Lemma 4.3.** $\mathcal{T}_{NonEmptyJoin}$ *can be filtered from* $\mathcal{T}_{Rels}$ *by an XQuery query.*

**Lemma 4.4.** *Let* $r, s : \text{Trees}_\tau \to \mathbb{N}$.
*If* $r(T) \cdot s(T) \in o\big(\frac{size(T)}{\log(size(T))}\big)$, *then* $\mathcal{T}_{EmptyJoin} \notin ST(r,s)$.

The proofs of these two lemmas can be found in Appendix F and G, respectively. From Lemma 4.4 and Lemma 4.3 we immediately obtain a lower bound on the worst-case data complexity for filtering relative to an XQuery query:

**Theorem 4.5.** *The tree-language* $\mathcal{T}_{EmptyJoin}$
*(a) can be filtered from* $\mathcal{T}_{Rels}$ *by an XQuery query,*
*(b) does not belong to the class* $ST(r,s)$, *whenever* $r, s : \text{Trees}_\tau \to \mathbb{N}$ *with*

$$r(T) \cdot s(T) \;\in\; o\Big(\frac{size(T)}{\log(size(T))}\Big).$$

*Remark 4.6.* Let us note that the above bound is "almost tight" in the following sense: The problem of deciding whether $A \bowtie_1 B = \emptyset$ and, in general, all *FO*-definable problems belong to $ST(1,n)$ – in its single scan of the external memory tape, the Turing machine simply copies the entire input on one of its internal memory tapes and then evaluates the *FO*-sentence by the straightforward LOGSPACE algorithm for *FO*-model-checking (cf. e.g. [1]).    $\square$

**Sorting.** By KeySort, we denote the problem of sorting a set $S$ of tuples $t = (K,V)$ consisting of a key $K$ and a value $V$ by their keys. Let $ST^-(r,s)$ denote the class of all problems in $ST(r,s)$ that can be solved without writing to the external memory tape. Then,

**Theorem 4.7.** *Let* $r, s : \mathbb{N} \to \mathbb{N}$. *If* KeySort *is in* $ST^-(r,s)$, *then computing the natural join* $A \bowtie B$ *of two finite relations* $A, B$ *is in*

$$ST^-\Big(r(n^2) + 2, s(n^2) + O(\log n) + O(\max_{t \in A \cup B} |t|)\Big).$$

A proof is given in Appendix H.

*Remark 4.8.* Given that the size of relations $A$ and $B$ is known (which is usually the case in practical database management systems DBMS), the algorithm given in the previous proof can do a merge-join without additional scans after the sort run and without a need to buffer more than one tuple. This is guaranteed even if both relations may contain many tuples with the same join key – in current implementations of the merge join in DBMS, this may lead to grass-roots swapping. The (substantial) *practical* drawback of the join algorithm of the proof of Theorem 4.7, however, is that much larger relations $A', B'$ need to be sorted: indeed $|A'| = |A| * |B|$.    $\square$

**Corollary 4.9.**

(a) *Let* $r, s : \mathbb{N} \to \mathbb{N}$ *such that* $r(n^2) \cdot \left(s(n^2) + \log n\right) \;\in\; o\!\left(\frac{n}{\log n}\right).$
    *Then,* KEYSORT $\notin ST^-(r, s)$.

(b) KEYSORT $\;\notin\; ST^-\!\left(o\!\left(\sqrt[5]{n}\right), O\!\left(\sqrt[5]{n}\right)\right).$

The proof is given in Appendix I.

It is straightforward to see that by using MergeSort, the sorting problem can be solved using $O(\log n)$ scans of external memory provided that *three* external memory tapes are available. (In [17], this logarithmic bound is shown to be tight, for arbitrarily many external tapes.) Corollary 4.9 gives an exponentially stronger lower bound for the case of a single external tape.

**A hierarchy based on the number of scans.**

**Theorem 4.10.** *For every fixed* $k \geqslant 1$,

$$ST\big(k,\, O((\log k) + \log n)\big) \;\cap\; NST\big(1,\, O(k \cdot \log n)\big) \quad \not\subseteq \quad ST\big(k-1, o\big(\tfrac{\sqrt{n}}{k^5 (\log n)^3}\big)\big).$$

The proof of this theorem, which can be found in Appendix J, is based on a result due to Duris, Galil and Schnitger [10]. Theorem 4.10 directly implies

**Corollary 4.11.** *For every fixed* $k \in \mathbb{N}$ *and all classes* $S$ *of functions from* $\mathbb{N}$ *to* $\mathbb{N}$ *such that* $O(\log n) \subseteq S \subseteq o\big(\tfrac{\sqrt{N}}{(\lg n)^3}\big)$ *we have* $ST(k, S) \subsetneqq ST(k+1, S)$.

*Remark 4.12.* On the other hand, of course, the hierarchy collapses if internal memory space is at least linear in the size of the input: For every $r : \mathbb{N} \to \mathbb{N}$ and for every $s : \mathbb{N} \to \mathbb{N}$ with $s(n) \in \Omega(n)$, we have

$$ST(r, s) \subseteq ST(1, n + s(n)) \quad \text{and} \quad ST(r, O(s(n))) = \text{DSPACE}(O(s(n))).$$

## 5  Tight bounds for filtering and query evaluation on trees

**Lower bound.** We need the following notation: We fix a set $\tau$ of tag names via $\tau := \{\text{ root, left, right, blank }\}$. Let $T_1$ be the $\tau$-tree from Figure 1. Note that $T_1$ has a unique leaf $v_1$ labeled with the tag name "left". For any arbitrary $\tau$-tree $T$ we let $T_1(T)$ be the $\tau$-tree rooted at $T_1$'s root and obtained by identifying node $v_1$ with the root of $T$ and giving the label "left" to this node. Now, for every $n \geqslant 2$ let $T_n$ be the $\tau$-tree inductively defined via $T_n := T_1(T_{n-1})$. It is straightforward to see that $T_n$ has exactly $2n$ leaves labeled "blank". Let $x_1, \ldots, x_n, y_n, \ldots, y_1$ denote these leaves, listed in *document order* (i.e., in the order obtained by a pre-order depth-first left-to-right traversal of $T_n$). For an illustration see Figure 2.

We let $\tau_{01} := \tau \cup \{0, 1\}$. For all sets $X, Y \subseteq \{1, \ldots, n\}$ let $T_n(X, Y)$ be the $\tau_{01}$-tree obtained from $T_n$ by replacing, for each $i \in \{1, \ldots, n\}$, (*) the label "blank" of leaf $x_i$ by the label 1 if $i \in X$, and by the label 0 otherwise and (*) the label "blank" of leaf $y_i$ by the label 1 if $i \in Y$, and by the label 0 otherwise.

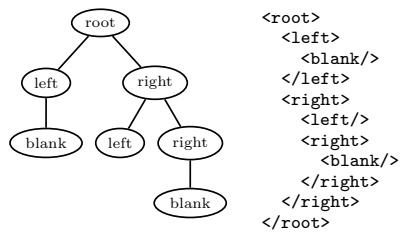

```
<root>
    <left>
        <blank/>
    </left>
    <right>
        <left/>
        <right>
            <blank/>
        </right>
    </right>
</root>
```

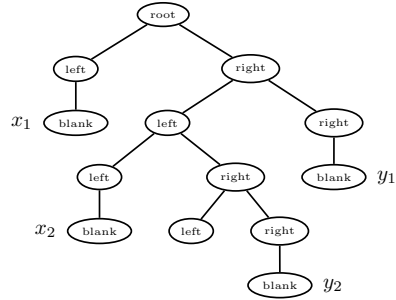

**Fig. 1.** A $\tau$-tree $T_1$ and its XML document $Doc(T_1) \in \Sigma_\tau^*$ with tag names $\tau := \{\text{root, left, right, blank}\}$.

**Fig. 2.** Tree $T_2$ and nodes $x_1, x_2, y_1, y_2$.

We let

$$\mathcal{T}_{Sets} := \big\{\, T_n(X,Y) \ : \ n \geqslant 1,\ X,Y \subseteq \{1,\ldots,n\} \,\big\},$$

$$\mathcal{T}_{Disj} := \big\{\, T_n(X,Y) \in \mathcal{T}_{Sets} \ : \ X \cap Y = \emptyset \,\big\},$$

$$\mathcal{T}_{NonDisj} := \big\{\, T_n(X,Y) \in \mathcal{T}_{Sets} \ : \ X \cap Y \neq \emptyset \,\big\}.$$

**Lemma 5.1.** *(a) There is a Core XPath query Q such that the following is true for all $\tau$-trees $T \in \mathcal{T}_{Sets}$: $Eval(Q, T) \neq \emptyset \iff T \in \mathcal{T}_{NonDisj}$.*
*(b) There is a FO-sentence $\varphi$ such that the following is true for all $\tau$-trees $T$: $T \models \varphi \iff T \in \mathcal{T}_{NonDisj}$.*

A proof of this lemma can be found in Appendix K.

**Lemma 5.2.** *Let $r, s : Trees_\tau \to \mathbb{N}$.*
*If $r(T) \cdot s(T) \in o(depth(T))$, then $\mathcal{T}_{NonDisj} \notin ST(r, s)$.*

The proof is similar to the proof of Lemma 4.4. It is given in Appendix L.

From Lemma 5.1 and Lemma 5.2 we directly obtain a lower bound on the worst-case data complexity of Core XPath filtering:

**Theorem 5.3.** *The tree-language $\mathcal{T}_{NonDisj}$*
*(a) can be filtered from $\mathcal{T}_{Sets}$ by a Core XPath query,*
*(b) is definable by an FO-sentence (and therefore, also definable by a Boolean MSO query and recognizable by a tree automaton), and*
*(c) does not belong to the class $ST(r, s)$, whenever*
    *$r, s : Trees_\tau \to \mathbb{N}$ with $r(T) \cdot s(T) \in o(depth(T))$.*

In the following subsection we match this lower bound with the corresponding upper bound.

**Upper bounds.** Recall that a tree-language $\mathcal{T} \subseteq Trees_\tau$ is definable by an *MSO*-sentence if, and only if, it is recognizable by an unranked tree automaton, respectively, if, and only if, the language $\{BinTree(T) : T \in \mathcal{T}\}$ of associated *binary* trees is recognizable by an ordinary (ranked) tree automaton (cf., e.g., [7, 9, 30]).

**Theorem 5.4 (implicit in [25, 29]).** *Let $\mathcal{T} \subseteq Trees_\tau$ be a tree-language. If $\mathcal{T}$ is definable by an MSO-sentence (or, equivalently, recognizable by a ranked or an unranked finite tree automaton), then $\mathcal{T} \in ST(1, depth(\cdot) + 1)$.*

A direct proof can be found in Appendix M.

Recall that every Core XPath query is equivalent to a unary *MSO* query. Thus a Core XPath filter can be phrased as an MSO sentence on trees. From the Theorems 5.4 and 5.3 we therefore immediately obtain a tight bound for Core XPath filtering:

**Corollary 5.5.** *(a) Filtering from the set of unranked trees with respect to every fixed Core XPath query $Q$ belongs to $ST\big(1, O(depth(\cdot))\big)$.*
*(b) There is a Core XPath query $Q$ such that, for all $r, s : Trees_\tau \to \mathbb{N}$ with $r(T) \cdot s(T) \in o\big(depth(T)\big)$, filtering w.r.t. $Q$ does not belong to $ST(r, s)$.*

Next, we provide an upper bound for the problem of computing the set $Eval(Q, T)$ of nodes in an input tree $T$ matching a unary MSO (or Core XPath) query $Q$. We first need to clarify what this means, because writing the subtree of each matching node onto the output tape requires a very large amount of internal memory (or a large number of head reversals on the external memory tape), and this gives us no appropriate characterization of the difficulty of the problem. We study the problem of computing, for each node matched by $Q$, its index in the tree, in the order in which they appear in the document $Doc(T)$. We distinguish between the case where these indexes are to be written to the output tape in ascending order and the case where they are to be output in descending (i.e., reverse) order.

**Theorem 5.6 (implicit in [26, 20]).** *For every unary MSO or Core XPath query $Q$, the problem of computing, for input trees $T$, the nodes in $Eval(Q, T)$*
*(a) in ascending order belongs to $ST(3, O(depth(\cdot)))$.*
*(b) in reverse order belongs to $ST(2, O(depth(\cdot)))$.*

A proof is given in Appendix N.

Note that this bound is tight: From Corollary 5.5(c) we know that, for some Core XPath query $Q$, not even *filtering* (i.e., checking whether $Eval(Q, T)$ is empty) is possible in $ST(r, s)$ if $r(T) \cdot s(T) \in o\big(depth(T)\big)$.

# References

1. S. Abiteboul, R. Hull, and V. Vianu. *Foundations of Databases.* Addison-Wesley, 1995.
2. G. Aggarwal, M. Datar, S. Rajagopalan, and M. Ruhl. On the streaming model augmented with a sorting primitive. In *Proc. FOCS'04*, pages 540–549.
3. N. Alon, Y. Matias, and M. Szegedy. The space complexity of approximating the frequency moments. *Journal of Computer and System Sciences*, 58:137–147, 1999.
4. A. Arasu, B. Babcock, T. Green, A. Gupta, and J. Widom. "Characterizing Memory Requirements for Queries over Continuous Data Streams". In *Proc. PODS'02*, pages 221–232, 2002.
5. B. Babcock, S. Babu, M. Datar, R. Motwani, and J. Widom. Models and issues in data stream systems. In *Proc. PODS'02*, pages 1–16.
6. Z. Bar-Yossef, M. Fontoura, and V. Josifovski. "On the Memory Requirements of XPath Evaluation over XML Streams". In *Proc. PODS'04*, pages 177–188, 2004.

7. A. Brüggemann-Klein, M. Murata, and D. Wood. "Regular Tree and Regular Hedge Languages over Non-ranked Alphabets: Version 1, April 3, 2001". Technical Report HKUST-TCSC-2001-05, Hong Kong Univ. of Science and Technology, 2001.

8. J.-E. Chen and C.-K. Yap. "Reversal Complexity". *SIAM J. Comput.*, **20**(4):622–638, Aug. 1991.

9. J. Doner. "Tree Acceptors and some of their Applications". *Journal of Computer and System Sciences*, **4**:406–451, 1970.

10. P. Duris, Z. Galil, and G. Schnitger. Lower bounds on communication complexity. *Information and Computation*, 73:1–22, 1987. Journal version of STOC'84 paper.

11. G. Gottlob and C. Koch. "Monadic Datalog and the Expressive Power of Web Information Extraction Languages". *Journal of the ACM*, **51**(1):74–113, 2004.

12. G. Gottlob, C. Koch, and R. Pichler. "Efficient Algorithms for Processing XPath Queries". In *Proc. VLDB 2002*, pages 95–106, Hong Kong, China, 2002.

13. G. Gottlob, C. Koch, and R. Pichler. "The Complexity of XPath Query Evaluation". In *Proc. PODS'03*, pages 179–190, San Diego, California, 2003.

14. G. Graefe. "Query Evaluation Techniques for Large Databases". *ACM Computing Surveys*, **25**(2):73–170, June 1993.

15. T. J. Green, G. Miklau, M. Onizuka, and D. Suciu. "Processing XML Streams with Deterministic Automata". In *Proc. ICDT'03*, 2003.

16. M. Grohe, C. Koch, and N. Schweikardt. Tight lower bounds for query processing on streaming and external memory data. In *Proc. ICALP*, 2005. To appear.

17. M. Grohe and N. Schweikardt. Lower bounds for sorting with few random accesses to external memory. In *Proc. PODS*, 2005. To appear.

18. M. Henzinger, P. Raghavan, and S. Rajagopalan. Computing on data streams. In *External memory algorithms*, volume 50, pages 107–118. DIMACS Series In Discrete Mathematics And Theoretical Computer Science, 1999.

19. J. E. Hopcroft and J. D. Ullman. Some results on tape-bounded Turing machines. *Journal of the ACM*, **16**(1):168–177, 1969.

20. C. Koch. "Efficient Processing of Expressive Node-Selecting Queries on XML Data in Secondary Storage: A Tree Automata-based Approach". In *Proc. VLDB 2003*, pages 249–260, 2003.

21. E. Kushilevitz and N. Nisan. *Communication Complexity*. Cambridge Univ. Press, 1997.

22. U. Meyer, P. Sanders, and J. Sibeyn, editors. *Algorithms for Memory Hierarchies*, volume 2832 of *Lecture Notes in Computer Science*. Springer-Verlag, 2003.

23. J. Munro and M. Paterson. Selection and sorting with limited storage. *Theoretical Computer Science*, 12:315–323, 1980.

24. S. Muthukrishnan. Data streams: algorithms and applications. In *Proc. 14th SODA*, pages 413–413, 2003.

25. A. Neumann and H. Seidl. "Locating Matches of Tree Patterns in Forests". In *Proc. 18th FSTTCS, LNCS 1530*, pages 134–145, 1998.

26. F. Neven and J. van den Bussche. "Expressiveness of Structured Document Query Languages Based on Attribute Grammars". *J. ACM*, **49**(1):56–100, Jan. 2002.

27. R. Ramakrishnan and J. Gehrke. *Database Management Systems*. McGraw-Hill, 2002.

28. L. Segoufin. "Typing and Querying XML Documents: Some Complexity Bounds". In *Proc. PODS'03*, pages 167–178, 2003.

29. L. Segoufin and V. Vianu. "Validating Streaming XML Documents". In *Proc. PODS'02*, 2002.

30. J. Thatcher and J. Wright. "Generalized Finite Automata Theory with an Application to a Decision Problem of Second-order Logic". *Math. Syst. Theory*, **2**(1):57–81, 1968.

31. P. van Emde Boas. "Machine Models and Simulations". In J. van Leeuwen, editor, *Handbook of Theoretical Computer Science*, volume 1, chapter 1, pages 1–66. Elsevier Science Publishers B.V., 1990.

32. J. Vitter. External memory algorithms and data structures: Dealing with massive data. *ACM Computing Surveys*, 33(2):209–271, June 2001.

33. World Wide Web Consortium. "XQuery 1.0 and XPath 2.0 Formal Semantics. W3C Working Draft (Aug. 16th 2002), 2002. `http://www.w3.org/XML/Query`.

34. A. Yao. Some complexity questions related to distributive computing. In *Proc. 11th STOC*, pages 209–213, 1979.

# APPENDIX

## A   Definitions of Trees and Streams

Let $\tau$ be a finite set. We will use $\tau$ as a set of *tag names*. We associate with $\tau$ a finite alphabet $\Sigma_\tau$ as follows: For each symbol $a \in \tau$, the alphabet $\Sigma_\tau$ contains (i) a symbol $\langle a \rangle$ (corresponding to the opening tag labeled $a$), (ii) a symbol $\langle /a \rangle$ (corresponding to the closing tag labeled $a$), and (iii) a symbol $\langle a/ \rangle$ (corresponding to the bachelor tag labeled $a$).

*Binary $\tau$-trees* are finite labeled ordered trees where each node has at most 2 children and is labeled with a symbol (i.e., tag name) in $\tau$.

*Unranked $\tau$-trees* are finite labeled ordered trees where each node may have an arbitrary number of children and is labeled with a symbol in $\tau$. We use *Trees$_\tau$* to denote the set of all unranked $\tau$-trees. An unranked $\tau$-tree $T$ can be represented by a binary tree $BinTree(T)$ in a straightforward way by using the *first-child / next-sibling* notation (cf., e.g., [11]).

The XML document $Doc(T)$ corresponding to an unranked $\tau$-tree $T$ can be viewed as a string over the alphabet $\Sigma_\tau$, cf. Figure 1 in the appendix. In particular, reading the string $Doc(T)$ from left to right corresponds to a *depth-first left-to-right traversal* of the tree $T$. For a set $\mathcal{T}$ of $\tau$-trees we write $Doc(\mathcal{T})$ for the string language $Doc(\mathcal{T}) := \{ Doc(T) : T \in \mathcal{T} \} \subseteq \Sigma_\tau^*$. We use $size(T)$ to denote the number of nodes in $T$, and we use $depth(T)$ to denote the maximum number of edges on a path from the root to one of $T$'s leaves.

## B   Definition of the Query Language Core XPath

XPath uses thirteen binary relations – called *axes* – for navigating in trees. We only introduce four of them in this paper, *Child* (the intuitive child relation; *Child(v, w)* iff $w$ is a child of $v$), *Parent* (its inverse), *Descendant* (the transitive closure of *Child*), and *Ancestor* (its inverse). In the following definition of Core XPath, we assume all 13 axes to be supported. (For a complete formal definition of (Core) XPath see [12].)

**Definition B.1.** Let $T$ be a tree. We define the syntax of *Core XPath* by the EBNF

| | |
|---|---|
| corexpath: | locationpath \| '/' locationpath |
| locationpath: | locationstep ('/' locationstep)* |
| locationstep: | $\chi$ '::' $P$ \| $\chi$ '::' $P$ '[' pred ']' |
| pred: | pred 'and' pred \| pred 'or' pred |
| | \| 'not(' pred ')' \| corexpath |
| | \| '(' pred ')' |

where "corexpath" is the start production, $\chi$ stands for an axis, and $P$ for a "node test", that is, a tag name from $\tau$ or "*", meaning "any node" $V^T$. We write $\mathcal{L}$(corexpath) to denote the language defined by the above EBNF for the symbol corexpath.

The semantics of Core XPath queries on trees $T$ is defined by two functions $\mathcal{S}$ and $\mathcal{E}$ (for Core XPath expressions and condition predicates, respectively):

$$\mathcal{S} \;:\; \mathcal{L}(\text{corexpath}) \to 2^{V^T \times V^T}$$

$$\mathcal{S}[\![\chi{::}P[e]]\!] := \{\langle x, y \rangle \mid \chi^T(x, y) \,\wedge$$
$$label^T(y) = P \;\wedge\; y \in \mathcal{E}[\![e]\!]\}$$

$$\mathcal{S}[\![/\pi]\!] := V^T \times \{x \mid \langle root^T, x \rangle \in \mathcal{S}[\![\pi]\!]\}$$

$$\mathcal{S}[\![\pi_1/\pi_2]\!] := \{\langle x, z \rangle \mid \exists y : \langle x, y \rangle \in \mathcal{S}[\![\pi_1]\!] \,\wedge$$
$$\langle y, z \rangle \in \mathcal{S}[\![\pi_2]\!]\}$$

$$\mathcal{E} \;:\; \mathcal{L}(pred) \to 2^{V^T}$$

$$\mathcal{E}[\![e_1 \text{ and } e_2]\!] := \mathcal{E}[\![e_1]\!] \cap \mathcal{E}[\![e_2]\!]$$

$$\mathcal{E}[\![e_1 \text{ or } e_2]\!] := \mathcal{E}[\![e_1]\!] \cup \mathcal{E}[\![e_2]\!]$$

$$\mathcal{E}[\![\text{not}(e)]\!] := V^T - \mathcal{E}[\![e]\!]$$

$$\mathcal{E}[\![\pi]\!] := \{x_0 \mid \exists x : \langle x_0, x \rangle \in \mathcal{S}[\![\pi]\!]\}$$

Here, $\pi$, $\pi_1$ and $\pi_2$ are location paths. Query $Q$ results in the set

$$Eval(Q, T) := \{y \mid \exists x : \langle x, y \rangle \in \mathcal{S}[\![Q]\!]\}.$$

$\square$

## C  Definition of the class $ST(r, s)$ for trees

Let $\tau$ be a set of tag names. Recall from section 2 that $Trees_\tau$ denotes the set of all unranked $\tau$-trees.

**Definition C.1 ($ST(r, s)$ for trees).** Let $r : Trees_\tau \to \mathbb{N}$ and $s : Trees_\tau \to \mathbb{N}$.

(a) A Turing machine $M$ is $(r, s)$-*bounded*, if every run $\rho$ of $M$ on an input string $Doc(T)$, for all $T \in Trees_\tau$ satisfies the following conditions:
   - $\rho$ is finite,
   - $\text{rev}(\rho) \leq r(T)$,
   - $\sum_{i=1}^{u} \text{space}(\rho, i) \leq s(T)$, where $u$ is the number of internal tapes of $M$.
(b) A tree-language $\mathcal{T} \subseteq Trees_\tau$ belongs to the class $ST(r, s)$, if there is a deterministic $(r, s)$-bounded Turing machine $M$ such that, for all $T \in Trees_\tau$, we have  $T \in \mathcal{T}$ if, and only if, $M$ accepts the string $Doc(T)$.    $\square$

## D  Proof of Lemma 4.1

*Proof of Lemma 4.1:*

For the sake of contradiction let us assume that there is an $(r, s)$-bounded Turing machine $M$ which accepts a string of the form $f_n(X)g_n(Y)$ if, and only if, $X \cap Y =$ $\emptyset$. Since $M$ is $(r, s)$-bounded, on an input string of length $N$, $M$'s internal memory tapes always have length $\leqslant s(N)$, and the external memory tape head can pass any particular external memory tape position $p$ for at most $r(N)$ times.

Let $n_0 \geqslant 1$ and let $n \geqslant n_0$ be chosen according the lemma's assumption. Let $X$ and $Y$ be arbitrary subsets of $\{1, \ldots, n\}$. We know that the string $f_n(X) g_n(Y)$ has length $\leqslant \lambda(n)$ and that $f_n(X) g_n(Y) \in L$ if, and only if, $X \cap Y = \emptyset$.

In particular, any internal memory tape configuration during a run of $M$ on $f_n(X) g_n(Y)$ can be represented by a bit-string of length $d \cdot s(\lambda(n))$, for a suitable constant $d$.

Let $Q$ denote $M$'s set of states. Using $M$, one obtains a communication protocol $\mathcal{P}_n$ that computes the disjointness function $Disj_n(\cdot, \cdot)$ as follows: Alice's input set $X \subseteq \{1, \ldots, n\}$ is represented by the string $f_n(X)$, whereas Bob's input set $Y$ is represented by the string $g_n(Y)$. Let $p := |f_n(X)|$. Alice starts the protocol by starting the Turing machine $M$ on input "$f_n(X) \cdots$" and letting it run until the first time $M$ tries to access the external memory tape position $p+1$. Then she sends the current state and internal memory tape configuration of $M$ to Bob. That is, she sends $\big(\log |Q| + d \cdot s(\lambda(n))\big)$ bits of information. Now, Bob has all the information needed to continue the execution of $M$ on input "$\cdots g_n(Y)$" until the first time $M$ tries to access the external memory tape position $p$. Then, Bob sends the current state and internal memory tape configuration of $M$ to Alice. Alice and Bob continue in this manner until the Turing machine $M$ stops, deciding whether or not $f_n(X) g_n(Y)$ belongs to $L$ and hence providing one of the players with the desired information whether or not $X \cap Y = \emptyset$.

Since $M$ passes the external memory tape position $p$ for at most $r(\lambda(n))$ times, the above protocol $\mathcal{P}_n$ computes the function $Disj_n(\cdot, \cdot)$ by exchanging at most

$$r(\lambda(n)) \cdot \big(\log |Q| + d \cdot s(\lambda(n))\big)$$

bits of information. However, since $r(\lambda(n)) \cdot s(\lambda(n)) \in o(n)$, we can find an $n_0 \in \mathbb{N}$ such that, for every $n \geqslant n_0$,

$$r(\lambda(n)) \cdot \big(\log |Q| + d \cdot s(\lambda(n))\big) \quad < \quad n.$$

Then, the above protocol $\mathcal{P}_n$ computes $Disj_n(\cdot, \cdot)$ with less than $n$ bits of communication, contradicting Theorem 2.2. $\qquad\square$

# E   Proof of Proposition 4.2

*Proof of Proposition 4.2:*
For every $n \in \mathbb{N}$ we choose functions $f_n, g_n : 2^{\{1, \ldots, n\}} \to \{0, 1, \#\}^*$ as follows: For every $X \subseteq \{1, \ldots, n\}$ let $f_n(X) := x\#$ and $g_n(X) := x$, where $x = x_1 \cdots x_n \in \{0, 1\}^n$ is the (unique) $n$-bit string with $S(x) = X$.

Then, for all $n \in \mathbb{N}$ and all $X, Y \subseteq \{1, \ldots, n\}$ we have

$$f_n(X) g_n(Y) \in L_{Disj} \iff X \cap Y = \emptyset,$$

and $|f_n(X) g_n(Y)| = 2n + 1 \leqslant 3n$.

Assuming that $r(n) \cdot s(n) \in o(n)$, we obtain from Lemma 4.1 that $L_{Disj} \notin ST(r, s)$. $\qquad\square$

# F   Proof of Lemma 4.3

*Proof of Lemma 4.3:*

We can choose the XQuery query $Q :=$

```
for $x in /rels/rel1/tuple/no1
  for $y in /rels/rel2/tuple/no1
    where deep-equal($x,$y) return <tuple/>
```

It is straightforward to see that for all finite $A, B \subseteq \mathbb{N}^2$, the result $Eval(Q, T(A, B))$ of $Q$ on the tree $T(A, B)$ returns one "tuple"-node for each tuple in $A \bowtie_1 B$. In particular, $Eval(Q, T(A, B))$ is *empty* if, and only if, $A \bowtie_1 B = \emptyset$.    □

# G   Proof of Lemma 4.4

*Proof of Lemma 4.4:*

We use Lemma 4.1. For finite $X, Y \subseteq \mathbb{N}$ let

$$A_X := \{ (i, 1) : i \in X \},$$
$$B_Y := \{ (i, 2) : i \in Y \}.$$

Obviously, $A_X \bowtie_1 B_Y = \emptyset$ if, and only if, $X \cap Y = \emptyset$.

For every $n \in \mathbb{N}$ we choose functions $f_n, g_n : 2^{\{1, \dots, n\}} \to \Sigma_\tau^*$ via

$$f_n(X) := \langle \text{rels} \rangle \; \langle \text{rel1} \rangle \; Doc(A_X) \; \langle /\text{rel1} \rangle$$
$$g_n(Y) := \langle \text{rel2} \rangle \; Doc(B_Y) \; \langle /\text{rel2} \rangle \; \langle /\text{rels} \rangle .$$

Then, for all $X, Y \subseteq \{1, \dots, n\}$, the string $f_n(X)g_n(Y) = Doc(T(A_X, B_Y))$ has length $O(n \cdot \log n)$, and

$$f_n(X)g_n(Y) \in Doc(\mathcal{T}_{EmptyJoin}) \iff X \cap Y = \emptyset.$$

From Lemma 4.1 we obtain for arbitrary $r', s' : \mathbb{N} \to \mathbb{N}$ with $r'(n \cdot \log n) \cdot s'(n \cdot \log n) \in o(n)$ that there is no $(r', s')$-bounded Turing machine which accepts exactly those strings of the form $f_n(X)g_n(Y)$ where $X \cap Y = \emptyset$. Noting that $size(T(A_X, B_Y)) \in$

$$\Theta\big( |Doc(T(A_X, B_Y))| \; = \; \Theta(|f_n(X)g_n(Y)|),$$

one then obtains for arbitrary $r, s : Trees_\tau \to \mathbb{N}$ with $r(T) \cdot s(T) \in o\big(\frac{size(T)}{\log(size(T))}\big)$ that $\mathcal{T}_{NonEmptyJoin} \notin ST(r, s)$.    □

# H   Proof of Theorem 4.7

*Proof Sketch of Theorem 4.7:*

We want to join two relations $A$ and $B$ by their common column(s) $K$. Let $n$

be the combined sizes of the binary representations of $A$ and $B$. Assume we can sort the tuples of a relation by a number of key columns using an $ST^-(r,s)$ algorithm.

Let us first briefly consider how a merge-join on sorted relations works (cf. e.g. [27]). Observe that on sorted relations $A$ and $B$ – and sort we can – a merge-join can compute a natural join $A \bowtie B$ using only buffer space for one tuple if there is a one-to-many relationship between $A$ and $B$; say, for each tuple $t \in A$, there is at most one tuple $t' \in B$ such that $t.K = t'.K$. The normal mode of operation for a merge-join is to have the two sorted relations on separate external memory tapes. But this is not necessary; we can simply add another column $RelId$ to both relations in which we store the name of the relation ("A" or "B") for each tuple, with "B" < "A" to get the single $B$-tuple before the possibly multiple matching $A$-tuples. Then we sort the union of the two modified relations on $(K, RelId)$ (with $K$ the more significant column). Now we can compute $A \bowtie B$ in a single forward scan of the sort, buffering a $B$-tuple whenever it is encountered (in a single slot, always replacing a previously inserted $B$-tuple) and comparing the $B$-tuple in the buffer with the $A$-tuples that follow.

| $A$ | | $B$ | | $B'$ | | | |
|---|---|---|---|---|---|---|---|
| $K$ | $V_A$ | $K$ | $V_B$ | $K$ | $V_B$ | $I_B$ | $RelId$ |
| $a$ | $x_1$ | $c$ | $y_1$ | $c$ | $y_1$ | 1 | "B" |
| $b$ | $x_2$ | $a$ | $y_2$ | $a$ | $y_2$ | 2 | "B" |
| $a$ | $x_3$ | $a$ | $y_3$ | $a$ | $y_3$ | 3 | "B" |

| $A'$ | | | | $\mathrm{sort}_{K,I_B,RelId}(A' \cup B')$ | | | | | | $A' \bowtie B'$ | | | |
|---|---|---|---|---|---|---|---|---|---|---|---|---|---|
| $K$ | $V_A$ | $I_B$ | $RelId$ | $K$ | $V_A$ | $V_B$ | $I_B$ | $RelId$ | | $K$ | $V_A$ | $I_B$ | $V_B$ |
| $a$ | $x_1$ | 1 | "A" | $a$ | $x_1$ | | 1 | "A" | | $a$ | $x_1$ | 2 | $y_2$ |
| $a$ | $x_1$ | 2 | "A" | $a$ | $x_3$ | | 1 | "A" | | $a$ | $x_1$ | 3 | $y_3$ |
| $a$ | $x_1$ | 3 | "A" | $a$ | | $y_2$ | 2 | "B" | | $a$ | $x_3$ | 2 | $y_2$ |
| $b$ | $x_2$ | 1 | "A" | $a$ | $x_1$ | | 2 | "A" | | $a$ | $x_3$ | 3 | $y_3$ |
| $b$ | $x_2$ | 2 | "A" | $a$ | $x_3$ | | 2 | "A" | | | | | |
| $b$ | $x_2$ | 3 | "A" | $a$ | | $y_3$ | 3 | "B" | | | | | |
| $a$ | $x_3$ | 1 | "A" | $a$ | $x_1$ | | 3 | "A" | | | | | |
| $a$ | $x_3$ | 2 | "A" | $a$ | $x_3$ | | 3 | "A" | | | | | |
| $a$ | $x_3$ | 3 | "A" | $b$ | $x_2$ | | 1 | "A" | | | | | |
| | | | | $b$ | $x_2$ | | 2 | "A" | | | | | |
| | | | | $b$ | $x_2$ | | 3 | "A" | | | | | |
| | | | | $c$ | | $y_1$ | 1 | "B" | | | | | |

**Fig. 3.** Example run of the join algorithm.

In general, there is no one-to-many relationship between $A$ and $B$, but we can use the following trick. Given an $ST^-(r,s)$ machine $M$ for sorting, we modify $M$ as follows:

– We add three $\log n$ size registers to the internal memory tape, which are called $sizeB$, $currentBidx$, and $iteratorA$. Moreover we need a register $tup$ for storing one tuple from either $A$ or $B$ (plus, strictly speaking, a log-sized register for storing the $I_B$ column that we will add to each $B$-tuple).

- Before we start with the usual operation of $M$, we make a first forward run over the external memory tape to count the tuples in $B$ and store $|B|$ in internal memory tape register $sizeB$. When this is done, we reverse and move back to the start of the external memory tape. (This accounts for two reversals of the *read* head on the external tape.)
- Then, we start the usual operation of $M$. However, whenever we see an $A$ tuple $(K : k, V : v)$ (which we may store on the internal memory tape) on the external memory tape, we simulate the reading of the $|B|$ tuples $(K : k, V_A : x, I_B : 1, RelId : \text{“A”}), \ldots, (K : k, V_A : x, I_B : |B|, RelId : \text{“A”})$. (We may use the *iteratorA* internal memory tape register to count up to $sizeB$ and copy an $A$-tuple into *tup* when we read it in order not to require further scans when we want to hand it over to $M$ multiple times.) That is, when $M$ asks to move the read head on to an $A$-tuple from the left, we show it the tuple with $I_B = 1$; when it asks to move on to the right, we show it the same tuple again with $I_B = 2$, and so on, until $I_B = |B|$. If $M$ asks to move on to the right, we really move on to the next tuple to the right. When we move to the left, we proceed analogously, but provide decreasing indexes $I_B$, starting with $|B|$.
  Whenever we see a $B$ tuple $(K : k, V_B : y)$ on the external memory tape, we simulate the reading of a tuple $(K : k, V_B : y, I_B : i, RelId : \text{“B”})$, where $i$ is the current value in the *currentBidx* register.
  Note that in order to simulate such a much larger tape requires the external tape to be read-only. But this is assured as $M$ is an $ST^-$-machine.
  Since we simulate a relation of size not greater than $n^2$, $M$ is in $ST^-(r(n^2), s(n^2))$.
- We sort on key $(K, I_B, RelId)$ (with decreasing significance from $K$ to $RelId$) of the $A$ and $B$ tuples, with sort order “A” $<$ “B” on the relation names in the $RelId$ column.
  Now observe that the simulated tuples define relations $A'$, $B'$ such that, in relational algebra,

  $$\pi_{K,V_A,V_B}(\pi_{K,V_A,V_B,I_B} A' \bowtie \pi_{K,V_A,V_B,I_B} B') = A \bowtie B,$$

  but now there is a guaranteed many-to-one relation between $A'$ and $B'$. An example of our construction is provided in Figure 3.
- While the tuples are produced in ascending sorted order by $M$, rather than writing them to the output immediately, we proceed as follows. Whenever we see a $B'$-tuple, we copy it into our *tup* register on the internal memory tape. Whenever we see an $A'$-tuple $t$ (and a $B'$-tuple has been seen before), we check whether $t.K = tup.K$ and $t.I_B = tup.I_B$. If there is a match, we produce a tuple $(K, V_A, V_B)$ and write it to the output.

It is easy to verify that this construction provides us with an $ST^-\big(r(n^2) + 2, s(n^2) + O(\log n + \max_{t \in A \cup B} |t|)\big)$ machine for joining two relations.     $\square$

## I   Proof of Corollary 4.9

*Proof Sketch of Corollary 4.9:*

*ad (a):* By contradiction. Suppose there is an $(r, s)$-bounded Turing machine which computes KEYSORT (i.e., writes the *sorted* version of the input to its write-only output tape), but which does not write to its external tape.

Then, by Theorem 4.7, we can compute the join of two input relations $A, B$ in

$$ST^-(r(n^2) + 2, s(n^2) + O(\log n) + O(\max_{t \in A \cup B} |t|)). \tag{1}$$

Now consider the problem JOIN$_{\log}$, which is defined as the restriction of the natural join problem to input relations $A, B$ where the size $|t|$ of (the binary representation of) each tuple $t \in A \cup B$ is at most *logarithmic* in the size of the binary representation of the input relations $A, B$. From (1) we then know that this problem JOIN$_{\log}$ must belong to

$$ST^-(r(n^2) + 2, s(n^2) + O(\log n)). \tag{2}$$

Let $r', s' : \mathbb{N} \to \mathbb{N}$ with $r'(n) := r(n^2) + 2$ and $s'(n) \in s(n^2) + O(\log n)$; thus JOIN$_{\log} \in ST(r', s')$.

From the corollary's assumption on the asymptotic size of $r, s$ we know that $r'(n) \cdot s'(n) \in o\left(\frac{n}{\log n}\right)$.

However, a variation of the proof of Lemma 4.4 shows that

$$\text{JOIN}_{\log} \quad \notin \quad ST(r', s'), \tag{3}$$

if $r', s' : \mathbb{N} \to \mathbb{N}$ with $r'(n) \cdot s'(n) \in o\left(\frac{n}{\log n}\right)$. (To prove this, one can use a variant of the *disjointness problem* $Disj_n(\cdot, \cdot)$ where it is known that each of the given sets $X, Y$ has at least $\frac{n}{3}$ elements – the communication complexity for deciding whether $X$ and $Y$ are disjoint then is $\geqslant \frac{n}{3}$.)

This completes the proof of (a).

*ad (b):* This is a direct consequence of (a), because

$$o((n^2)^{\frac{1}{5}}) \cdot \left(O((n^2)^{\frac{1}{5}}) + O(\log n)\right) \quad \subseteq \quad o(n^{\frac{2}{5}}) \cdot O(n^{\frac{2}{5}}) \quad \subseteq \quad o(n^{\frac{4}{5}}) \quad \subseteq \quad o(\frac{n}{\log n})$$

$$\square$$

## J    Proof of Theorem 4.10

Duris, Galil, and Schnitger [10] prove an exponential gap between $k$- and $k{+}1$-round communication complexity. They consider functions $f : \{0, \dots, 2^m{-}1\} \to \{0, \dots, 2^m{-}1\}$, encoded as list of binary representations of the values $f(0)$, $f(1)$, ..., $f(2^m{-}1)$, and prove a lower bound on the $k$-round communication complexity of the language $L_{k+1}$, consisting of the encodings of functions $f$ where

$$\underbrace{f(f(\cdots f(f(0)) \cdots))}_{k+2} \quad = \quad 2^m{-}1 .$$

The precise definition of $L_{k+1}$ is as follows:

**Definition J.1.** For every $k \in \mathbb{N}$, let  $L_{k+1} :=$

$$\{\, w_0 w_1 \cdots w_{2^m-1} \ : \ m \in \mathbb{N},\ w_i \in \{0,1\}^m,\ \text{and ex. } j_1, \ldots, j_{k+1}$$
$$\text{such that } w_0 = j_1, w_{j_i} = j_{i+1},\ w_{j_{k+1}} = 2^m - 1 \,\}. \quad \square$$

**Theorem J.2 (Duris, Galil, Schnitger [10]).** *For every $k \geqslant 1$, the following is true for all sufficiently large $n \in \mathbb{N}$:*

$$comm\text{-}compl_{k+1}(F_{k+1,n}) \leqslant (k+1) \cdot \log n, \quad but$$
$$comm\text{-}compl_k(F_{k+1,n}) \geqslant \frac{\sqrt{n}}{36 \cdot k^4 \cdot (\log n)^3},$$

*where the function $F_{k+1,n} : \{0,1\}^n \times \{0,1\}^n \to \{0,1\}$ is given via*

$$F_{k+1,n}(x,y) := \begin{cases} 1, & \text{if } xy \in L_{k+1} \\ 0, & \text{otherwise.} \end{cases}$$

In fact, Duris et al. [10] prove an even stronger result, namely that their lower bound applies for all $k$-round protocols, even if communication complexity is measured as the minimum complexity over all arbitrary partitions of the input bits into two parts of equal size.

*Proof of Theorem 4.10:*
We use Theorem J.2 and let  $L'_{k+1} :=$

$$\{\, 1^m \# w_0 \cdots w_{2^m-1} \ : \ m \in \mathbb{N},\ w_i \in \{0,1\}^m,\ w_0 \cdots w_{2^m-1} \in L_{k+1} \,\},$$

where $L_{k+1}$ is the language fixed in Definition J.1.

From the definition of $L_{k+1}$ it is straightforward to see that $L'_{k+1}$ belongs to $ST(k+1, O((\log k) + (\log n)))$ – the Turing machine just has to store the current index $i \in \{0, \ldots, k+1\}$ and the corresponding string $w_{j_i}$ on its internal tapes and move the external tape head to the block of index $j_{i+1} := w_{j_i}$. To recognize $L'_{k+1}$, this requires at most $k$ changes of the direction of the external tape head and internal space $O((\log k) + (\log n))$.

A *nondeterministic* Turing machine with internal space $\Omega(k \cdot \log n)$ does not even need a single reversal of the external tape head – it can simply guess the strings $w_{j_1}, \ldots, w_{j_{k+1}}$ on one of its internal tapes and verify their "correctness" while scanning the external tape from left to right.

Assume, for the sake of contradiction, that $L'_{k+1} \in ST\big(k, o\big(\frac{\sqrt{n}}{k^5(\log n)^3}\big)\big)$ via a Turing machine $M$ that is $(k,s)$-bounded, for some function $s : \mathbb{N} \to \mathbb{N}$ with $s(n) \in o\big(\frac{\sqrt{n}}{k^5(\log n)^3}\big)$. Then, in the same way as in the proof of Lemma 4.1, $M$ leads to a $k$-round protocol $\mathcal{P}_n$, for all $n \in \mathbb{N}$, that computes the function $F_{k+1,n}$ from Theorem J.2 and has cost at most $d \cdot k \cdot s(n)$, for a suitable constant $d$ (depending on $M$, but not on $k$ or $n$). Since $s(n) \in o\big(\frac{\sqrt{n}}{k^5(\log n)^3}\big)$, we can find sufficiently large $n$ such that $d \cdot s(n) < \frac{\sqrt{n}}{36 k^5 (\log n)^3}$. Consequently, for such $n$ we have $comm\text{-}compl_k(F_{k+1,n}) \leqslant d \cdot k \cdot s(n) < \frac{\sqrt{n}}{36 k^4 (\log n)^3}$, contradicting Theorem J.2.
$\square$

# K   Proof of Lemma 5.1

*Proof of Lemma 5.1:*

*ad (a):* We can choose $Q :=$

$$/descendant::*[child::right/child::right/child::1\ ]/child::left/child::1$$

which selects all nodes $x$ that are labeled 1 and for which there exists a node $z$ such that

(i) there exists a child $z'$ of $z$ which is labeled "left" such that $x$ is a child of $z'$,

(ii) there exists a child $z''$ of $z$ which is labeled "right" and has a child $z'''$ labeled "right" that has a child labeled 1.

It is straightforward to check that for all $T(X,Y) \in \mathcal{T}_{Sets}$ we have that $Q(T(X,Y))$ consists of exactly those nodes $x_i$ for which both, $x_i$ and $y_i$ are labeled 1. I.e., $Q(T(X,Y)) = \{x_i : i \in X \cap Y\}$.

*ad (b):* The above query $Q$ can be translated in a straightforward way into an *FO*-formula $\psi(x)$. The desired *FO*-sentence $\varphi$ is chosen as $\varphi := \chi \wedge \exists x\ \psi(x)$, where $\chi$ is a suitable *FO*-sentence expressing that the underlying tree has the correct shape.                                                                    □

# L   Proof of Lemma 5.2

*Proof of Lemma 5.2:*

We use Lemma 4.1. For every $n \in \mathbb{N}$ let $p_n$ denote the position in the string $Doc(T_n)$ that carries the unique leaf of $T_n$ carrying the label "left".

For every $n \in \mathbb{N}$ we choose functions $f_n, g_n : 2^{\{1,\dots,n\}} \to \Sigma^*_{\tau_{01}}$ as follows: For every $X \subseteq \{1,\dots,n\}$ let $f_n(X)$ be the prefix of $Doc(T_n(X,Y))$ up to position $p_n$, and let $g_n(Y)$ be the suffix of $Doc(T_n(X,Y))$ starting at position $p_n + 1$. Then, the string $f_n(X)g_n(Y) = Doc(T_n(X,Y))$ has length $10n+1 \leqslant 11 \cdot n$, and

$$f_n(X)g_n(Y) \in Doc(\mathcal{T}_{Disj}) \iff X \cap Y = \emptyset.$$

From Lemma 4.1 we obtain for arbitrary $r', s' : \mathbb{N} \to \mathbb{N}$ with $r'(n) \cdot s'(n) \in o(n)$ that there is no $(r', s')$-bounded Turing machine which accepts exactly those strings of the form $f_n(X)g_n(Y)$ where $X \cap Y = \emptyset$.

Noting that $depth(T_n(X,Y)) = 2n+2 \in \Theta(|Doc(T_n(X,Y))|) = \Theta(|f_n(X)g_n(Y)|)$ one then obtains for arbitrary $r, s : Trees_\tau \to \mathbb{N}$ with $r(T) \cdot s(T) \in o(depth(T))$ that $\mathcal{T}_{NonDisj} \notin ST(r,s)$.                                                                    □

# M   Proof of Theorem 5.4

*Proof Sketch of Theorem 5.4:*

We proceed in two steps.

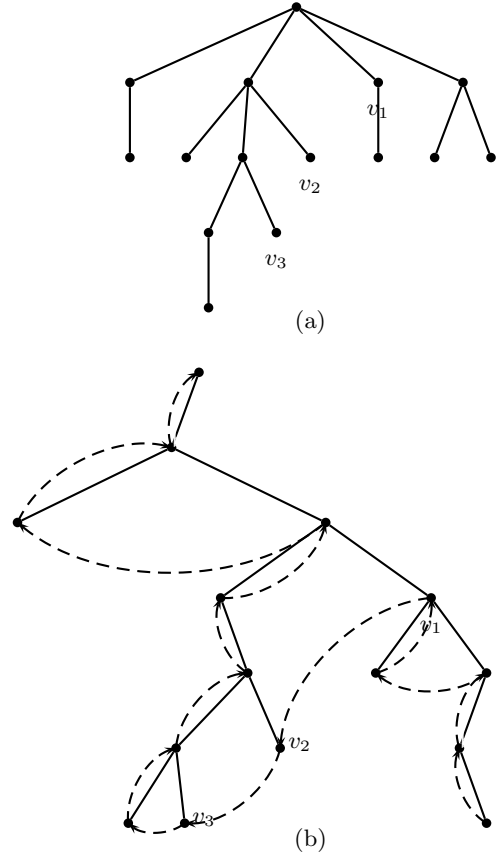

(a)

(b)

**Fig. 4.** Unranked ordered tree (a) and corresponding binary (first-child, next-sibling)-tree (b) with traversal order for the evaluation of a bottom-up automaton.

*Step 1:* We first show that $\mathcal{T} \in ST(1, depth(\cdot)+1)$.

Let $\mathcal{B}$ be a bottom-up binary tree automaton which accepts exactly the binary trees $BinTree(T)$ for $T \in \mathcal{T}$. For simplicity, we assume a single transition function $\delta^{\mathcal{B}} : \Sigma \times (Q \cup \{\bot\}) \times (Q \cup \{\bot\}) \to Q$.

We may assume that the input XML document $Doc(T)$ consists of a well-formed sequence of opening $\langle a \rangle$ and closing tags $\langle /a \rangle$, for tag symbols $a \in \tau$.[5] We evaluate $\mathcal{B}$ as follows, using a stack of states of the automaton $\mathcal{B}$. First we scan the input to the end. Then we reverse and scan it backwards. While scanning backwards, we do the following for each symbol $s$ seen:

**if** $s$ is a closing tag **then**
**begin**
    **if** there was no previous symbol or

---

[5] Non well-formed input can easily be detected by putting opening tags on the stack that we maintain in the algorithm below.

> it was a closing tag **then**
> > push($\bot$);
> **end**
> **else if** $s$ is an opening tag $\langle a \rangle$ **then**
> **begin**
> > **if** the previous symbol was a closing tag **then**
> > > $q_1 := \bot$;
> > **else**
> > > $q_1 := \text{pop}()$;
> >
> > $q_2 := \text{pop}()$;
> > $q := \delta^{\mathcal{B}}(a, q1, q2)$;
> > push($q$);
> **end**

Consider the example run of Figure 4. Just after having processed the opening tag of node $v_3$, the stack contains the symbols $\bot, \rho^{\mathcal{B}}(v_1), \rho^{\mathcal{B}}(v_2), \rho^{\mathcal{B}}(v_3)$ (where the final symbol is the top of the stack, and $\rho^{\mathcal{B}}(v)$ denotes the state assigned to node $v$ by the run $\rho$ of the tree automaton $\mathcal{B}$).

It is easy to verify that whenever we are at a node $v$ at depth $d$ in the unranked tree $T$ (i.e., between the opening and the closing tag of $v$, and not between the opening and closing tag of a descendant of $v$), there are $d+1$ items on the stack. Thus the depth of the stack never exceeds $depth(T)+1$. Since every stack entry consist of a single symbol, the space consumption of the internal memory tape is bounded by $depth(T)+1$.

On termination of this $ST\big(1, O\big(depth(\cdot)\big)\big)$ algorithm, the stack will contain precisely one symbol.

It is not difficult to verify that $\mathcal{B}$ accepts $BinTree(T)$ if and only if after processing the final (and thus leftmost) symbol of the input, the top of the stack holds a final state of $\mathcal{B}$.

*Step 2: From $ST(1, depth(\cdot)+1)$ to $ST(0, depth(\cdot)+1)$.*

Just as a binary bottom-up tree automaton on the (first-child, next-sibling) representation of (unranked) $\tau$-trees can be computed, so can a binary tree automaton $\mathcal{B}$ be computed that works on a (last-child, next-sibling) binary tree representation.

We can evaluate $\mathcal{B}$ in one single forward scan of the input by taking the algorithm of *Step 1*, and exchanging every occurrence of "opening tag" by "closing tag" and vice-versa. Now we need only one forward scan to check whether $\mathcal{B}$ accepts.

Altogether, the proof of Theorem 5.4 is complete.                    □

## N    Proof of Theorem 5.6

*Proof Sketch of Theorem 5.6:*

(a) In [20], a technique for evaluating unary MSO queries in two scans of the data is described. The first scan is a backward bottom-up tree automaton scan that writes the states computed for the nodes visited to an output tape that the second scan, a forward scan during which a top-down deterministic tree automaton is evaluated, reads.

Let $\mathcal{A}$ and $\mathcal{B}$ be a pair of a bottom-up and a top-down automaton for evaluating XPath expression $\pi$ in this way.

After scanning to the end of the input, we perform a backward scan during which we compute the run of $\mathcal{A}$ as described in the first part of the proof of Proposition 5.4. Here we always replace the opening tag $\langle a \rangle$ of node $v$ on the tape by a symbol $\langle a\ q = \rho^{\mathcal{A}}(v)\rangle$. (This is again a single tape symbol as both $\Sigma$ and the state set $Q^{\mathcal{A}}$ are fixed.) At the end of this run, we have, for each node $v$, the state $\rho^{\mathcal{A}}(v)$ computed by the run of $\mathcal{A}$ attached to it. Note that in the algorithm of the proof of Proposition 5.4, $\rho^{\mathcal{A}}(v)$ always gets available when the head on the external memory tape is on the position of the opening tag of node $v$, so we need no further buffer space besides the space occupied for the stack. Then we perform a third scan, a forward scan during which we compute the run of $\mathcal{B}$. $\mathcal{B}$ is a deterministic top-down tree automaton and the state of a node depends only on its label and the state of its parent. As $\mathcal{B}$ runs on the (first-child,next-sibling) presentation of unranked trees, we have always $\rho^{\mathcal{B}}(v)$ available as soon as we read the opening tag of node $v$. According to the construction of [20], the state $\rho^{\mathcal{B}}(v)$ indicates whether $v$ is in the query result. We maintain a counter (initialized with 0) and during the scan, whenever we see an opening tag we increment it by one. Thus, whenever we decide that a node is part of the output, we write the current value of the counter – which is the index of the node in document order – to the output tape. This gives us the nodes matching the query in ascending order.

(b) Using the same ideas as in the proof of Theorem 5.4 (changing the automata from running on (first-child, next-sibling) to (last-child, nextsibling)-trees), we can compute the indexes of nodes matching a unary MSO query in *reverse order* (i.e., we output the node indexes while traversing the data backwards). □

*Remark N.1.* The proof of Theorem 5.6 requires (i) to scan the external memory tape both forward and backward, and (ii) to store states of the bottom-up automaton used in the proof construction of Theorem 5.4 on the external tape. If the query is considered fixed (data complexity), states are constant-size and can replace symbols of the input; but this means that we need to allocate space enough to store a state into each tape position of the input. The results of [20] only readily yield automata $\mathcal{A}$ whose state space is of size doubly exponential in the size of the given query (in the query language of the framework, monadic datalog). If we want to use this technique, we need Turing machines whose external tape alphabet is of size doubly exponential in the size of the given query.